\documentclass[aps,pre,reprint,superscriptaddress,amsmath,amssymb,floatfix]{revtex4-2}
\usepackage{graphicx}
\usepackage{bm}
\usepackage{booktabs}
\usepackage{xcolor}
\usepackage[percent]{overpic}
\usepackage[colorlinks=true,citecolor=blue,linkcolor=blue,urlcolor=blue]{hyperref}
\graphicspath{{figures/}}

\newcommand{\Fx}{F^{\mathrm{ext}}}
\newcommand{\Tx}{\tau^{\mathrm{ext}}}
\newcommand{\zeff}{z_{\mathrm{eff}}}
\newcommand{\Pe}{\mathrm{Pe}}
\newcommand{\dbh}{d_{\mathrm{BH}}}

\begin{document}

\title{Activated coarsening of motility-induced phase separation in random environments}

\author{Parameshwaran A}
\affiliation{Department of Physics, Vellore Institute of Technology, Vellore, Tamil Nadu 632014, India}
\author{Bhaskar Sen Gupta}
\email{bhaskar@iiserbpr.ac.in}
\affiliation{Department of Physical Sciences, Indian Institute of Science Education and Research Berhampur, Ganjam, Odisha 760003, India}

\date{\today}

\begin{abstract}
Self-propelled particles that only repel one another can still separate into a dense and a dilute phase, a process known as motility-induced phase separation (MIPS). In a clean system the domains coarsen like any conserved mixture: domains grow as $\ell(t)\sim t^{1/3}$. Living and synthetic swimmers, however, move through rough, porous, or patterned surroundings. Using large-scale simulations of active Brownian particles, we ask how frozen heterogeneity changes MIPS coarsening. We introduce disorder in two ways, as a quenched random force field acting on particle positions and as a quenched random torque field acting on their orientations. Both destroy Lifshitz-Slyozov growth. A transient power law with a disorder-dependent exponent gives way to activated dynamics in which the effective dynamic exponent grows without bound and domains grow at most logarithmically. The crossover obeys the scaling form known from the random-field Ising model, and the domain morphology depends on disorder strength, so superuniversality fails. The two kinds of disorder act through different mechanisms. Random forces trap particles in the sinks of a random drift field once the drift beats self-propulsion. Random torques act as a sign-random, quenched chirality that erodes persistence and lowers the local Péclet number toward its critical value. Our results show that quenched disorder is a relevant perturbation for the kinetics of active phase separation and connect MIPS coarsening to the physics of pinned interfaces in disordered magnets.
\end{abstract}

\maketitle

%===========================================================
\section{Introduction}
\label{sec:intro}
%===========================================================

Active matter consists of units that take up energy from their surroundings and turn it into persistent motion~\cite{Marchetti2013,Bechinger2016}. Because this energy input happens at the level of each particle, active systems are driven out of equilibrium from the smallest scale up, and they show collective behavior that has no counterpart in thermal equilibrium. Examples range from bacterial suspensions and cell tissues to vibrated grains and synthetic colloidal swimmers.

One of the most robust of these collective effects is motility-induced phase separation (MIPS)~\cite{Tailleur2008,Fily2012,Redner2013,Cates2015}. Self-propelled particles that interact only through steric repulsion can nonetheless separate into a dense, nearly jammed phase and a dilute gas. The mechanism is a simple positive feedback. Particles slow down when they collide, so they accumulate where the density is high, which in turn slows them further. Above a threshold activity this feedback makes the homogeneous state unstable. MIPS has been observed in suspensions of light-activated and catalytic Janus colloids~\cite{Buttinoni2013,Palacci2013}, and minimal models such as active Brownian particles (ABPs) have mapped out its phase diagram in considerable detail, including a critical point~\cite{Siebert2018}.

In contrast with its unusual origin, the kinetics of MIPS turn out to be surprisingly ordinary. After a quench into the two-phase region, dense clusters nucleate or form by spinodal decomposition, merge, and grow. Simulations of ABPs and of scalar field theories such as Active Model B find that the characteristic domain size grows as $\ell(t)\sim t^{1/3}$~\cite{Stenhammar2014,Wittkowski2014,Caporusso2020,Dittrich2023}. This is the Lifshitz-Slyozov (LS) law~\cite{Lifshitz1961} for diffusive coarsening with a conserved order parameter, the same law that governs a binary alloy or a phase-separating fluid mixture at early times~\cite{Binder1977,Puri2009,Bhattacharyya2021,Bray1994}. As far as coarsening is concerned, activity changes prefactors and time scales, but not the universality class.

All of this refers to an idealized, perfectly uniform environment. Real active particles almost never move through one. Bacteria swim through soil, mucus, and tissue, where they are trapped and released by pores and obstacles~\cite{Bhattacharjee2019}. Synthetic swimmers move over rough or chemically patterned substrates, and experiments deliberately expose them to random optical potentials such as speckle fields~\cite{Volpe2014,Pince2016} or to spatially varying illumination that controls their activity~\cite{Lozano2016}. Quenched heterogeneity of this kind has a strong effect on single active particles, leading to trapping, subdiffusion, and optimal-noise effects~\cite{Chepizhko2013a,Chepizhko2013b,Reichhardt2014}. It distorts or destroys polar flocks~\cite{Morin2017,Preethi2026,Parameshwaran2026,Toner2018,Das2018} and gives rise to pinned, clogged, and flowing phases of active and driven assemblies~\cite{Sandor2017,Reichhardt2017}. For MIPS itself, Ro \emph{et al.}\ showed that quenched random potentials and torques destroy bulk phase separation below four dimensions beyond an Imry-Ma length and generate long-range density correlations~\cite{Ro2021,Imry1975}. Related work showed that even disorder confined to the boundaries is enough to destroy bulk phase separation~\cite{BenDor2022}.

These results concern the steady state. The question we address here is dynamical: how does quenched disorder change the \emph{coarsening} of MIPS? Equilibrium physics offers a clear reference. In random-field and random-bond magnets, disorder pins domain walls. The free-energy barriers that a domain must overcome to grow increase with its size, growth becomes thermally activated, and algebraic coarsening gives way to logarithmically slow growth~\cite{Huse1985,Fisher1988}. Numerical studies have found that at accessible times this often shows up as an apparent power law with a disorder-dependent exponent~\cite{Paul2004}, and Corberi \emph{et al.}\ proposed a scaling description of the crossover from this transient regime to activated growth in the random-field Ising model~\cite{Corberi2012}. A further question is whether the scaling functions depend on disorder at all. The superuniversality hypothesis states that they do not, and that disorder only changes the growth law of $\ell(t)$~\cite{Fisher1988}; it holds in some disordered magnets and fails in others~\cite{Paul2004,Corberi2012}.

It is not obvious that any of this carries over to active matter. Active particles never stop moving and inject energy locally and persistently. One might expect this drive to act like a high effective temperature that simply smooths the disorder away. On the other hand, the persistence that drives MIPS is itself fragile, and disorder can couple to it directly. Moreover, disorder in an active system can act on two quite different degrees of freedom. It can act on positions, as an external force that competes with self-propulsion, or on orientations, as a torque that competes with rotational diffusion and persistence. There is no equilibrium analogue of the second case.

In this paper we study both cases in detail using large-scale Brownian dynamics simulations of ABPs in two dimensions. We introduce a quenched random force field and, separately, a quenched random torque field, and we follow the coarsening of MIPS through the real-space morphology, the single-particle dynamics, the equal-time correlation function, the structure factor, and the growth of the domain size. Our main findings are as follows.

(i) Both kinds of disorder slow down coarsening and destroy the LS law. The effective dynamic exponent $\zeff(\ell)=d\ln t/d\ln\ell$ passes through a disorder-dependent plateau and then rises without bound.

(ii) This crossover follows the scaling form of Corberi \emph{et al.}~\cite{Corberi2012}. It is equivalent to an Arrhenius law with barriers that grow with domain size and implies logarithmically slow growth, $\ell\sim\lambda(\ln t)^{1/\varphi}$, with $\phi\approx 2.2$ for random forces and $\phi \approx 1.3$ for random torques, and a crossover length $\lambda$ that decreases with disorder.
%\textcolor{green}{ }

(iii) Coarsening remains self-similar at fixed disorder, but the scaling functions depend on disorder, so superuniversality fails. The conserved-order-parameter dip in the correlation function and the Porod tail of the structure factor disappear.

(iv) The two kinds of disorder act through different mechanisms, which we identify and quantify. Random forces trap particles in the sinks of a non-conservative random drift and act as a random field on the density. Random torques act as a quenched, sign-random chirality that lowers the local P\'eclet number. As a result, torques arrest coarsening at amplitudes several times smaller than forces.

The paper is organized as follows. Section~\ref{sec:model} introduces the model, the disorder fields, and the observables. Sections~\ref{sec:force} and \ref{sec:torque} present the results for random forces and random torques. Section~\ref{sec:mechanism} discusses the microscopic mechanisms. Section~\ref{sec:discussion} places the results in a broader context and discusses their limitations, and Sec.~\ref{sec:conclusion} summarizes. Two Appendices contain derivations.

%===========================================================
\section{Model and methods}
\label{sec:model}
%===========================================================

\subsection{Active Brownian particles}

We consider $N$ active Brownian particles in a two-dimensional square box of side $L_x=L_y=L$ with periodic boundary conditions. Particle $i$ has position $\mathbf r_i$ and orientation $\hat{\mathbf n}_i=(\cos\theta_i,\sin\theta_i)$, along which it self-propels with constant speed $v_0$. The overdamped equations of motion are~\cite{Stenhammar2014,Siebert2018,Dittrich2023}
\begin{align}
\dot{\mathbf r}_i &= v_0\hat{\mathbf n}_i-\frac{D_t}{k_BT}\left[\nabla_{\mathbf r_i}U-\mathbf f(\mathbf r_i)\right]+\sqrt{2D_t}\,\bm\xi_i(t),\label{eq:eom_r}\\
\dot\theta_i &= \frac{D_r}{k_BT}\,\Gamma(\mathbf r_i)+\sqrt{2D_r}\,\eta_i(t),\label{eq:eom_theta}
\end{align}
where $D_t$ and $D_r$ are the translational and rotational diffusion coefficients, and $\bm\xi_i$ and $\eta_i$ are Gaussian white noises with zero mean and correlations $\langle\xi_i^\alpha(t)\xi_j^\beta(t')\rangle=\delta_{ij}\delta^{\alpha\beta}\delta(t-t')$ and $\langle\eta_i(t)\eta_j(t')\rangle=\delta_{ij}\delta(t-t')$. The fields $\mathbf f(\mathbf r)$ and $\Gamma(\mathbf r)$ are the quenched random force and torque described below.

The interaction energy $U$ is a sum of Weeks-Chandler-Andersen (WCA) pair potentials~\cite{Weeks1971},
\begin{equation}
U_{\mathrm{WCA}}(r)=
\begin{cases}
4\varepsilon\left[\left(\dfrac{\sigma}{r}\right)^{12}-\left(\dfrac{\sigma}{r}\right)^{6}\right]+\varepsilon, & r<2^{1/6}\sigma,\\[2mm]
0, & r\ge2^{1/6}\sigma,
\end{cases}
\label{eq:wca}
\end{equation}
with $\sigma=1$ and $\varepsilon=100$. The large value of $\varepsilon$ makes the repulsion steep enough that particles hardly overlap even at the high swim speeds used here.

\subsection{Parameters}

We work in units with $\sigma=k_BT=D_t=1$. The effective particle diameter is given by the Barker-Henderson expression~\cite{Barker1967}
\begin{equation}
\dbh=\int_0^\infty\left[1-e^{-U_{\mathrm{WCA}}(r)/k_BT}\right]dr,
\label{eq:dbh}
\end{equation}
which yields $\dbh=1.10688$. The rotational diffusion coefficient is set to $D_r=3D_t/\dbh^2\simeq2.449$, the value for a sphere in a viscous fluid, and the area fraction is $\phi_A=N\pi\dbh^2/(4L^2)=0.597$. With $L=256$ this corresponds to $N\simeq4.07\times10^4$ particles. The activity is measured by the P\'eclet number
\begin{equation}
\Pe=\frac{3v_0}{D_r\dbh},
\label{eq:pe}
\end{equation}
the ratio of the persistence length $v_0/D_r$ to the particle size. We fix $v_0=125$, which gives $\Pe\simeq138.4$. For this model the MIPS critical point lies at $\Pe_c\simeq40$ and an area fraction close to the one used here~\cite{Siebert2018}, so the clean system is deep in the two-phase region.

\subsection{Quenched random force and torque fields}
\label{sec:disorder}

The disorder lives on a square lattice of unit spacing $a=\sigma$ that covers the simulation box. On each lattice cell we draw, once and for all, a force vector $\mathbf F^{ext}=(f_x,f_y)$ whose two components are independent and uniformly distributed in $[-\Fx,\Fx]$, and a scalar torque $\Gamma$ uniformly distributed in $[-\Tx,\Tx]$~\cite{Ahmad2014}. A particle feels the force and torque of the cell that contains it. The fields are frozen in time and uncorrelated from cell to cell, so their correlation length is one particle diameter. We study the two kinds of disorder separately: either $\Tx=0$ and $\Fx>0$, or $\Fx=0$ and $\Tx>0$.

Before presenting results it is worth being precise about the nature of these fields, because it matters for their interpretation.

\emph{The force field is not an energy landscape.} Because the two components of $\mathbf f$ are independent on every cell, $\mathbf f$ is not the gradient of any potential. By the Helmholtz decomposition it can be written as
\begin{equation}
\mathbf f=-\nabla V+\nabla\times(A\hat{\mathbf z}),
\label{eq:helmholtz}
\end{equation}
and both the gradient part (with potential $V$) and the rotational part (with stream function $A$) are present with comparable strength. Particles therefore do not explore a potential-energy landscape. They are carried by a frozen random drift $D_t\mathbf f/k_BT$ that has sources and sinks (from the divergence of $\mathbf f$) as well as vortices (from its curl). With $D_t=k_BT=1$, the drift velocity equals $\mathbf f$ and can be compared directly with $v_0$. We therefore quote the force disorder by the ratio $\Fx/v_0 =\chi$.

\emph{The torque field is a local angular velocity.} By Eq.~\eqref{eq:eom_theta}, a torque $\Gamma$ rotates the orientation at the angular velocity $\omega=D_r\Gamma/k_BT$. A particle that stays in one cell, as particles inside a dense cluster do, spins at a fixed rate, like a chiral (circle) swimmer. A particle that moves through the system picks up a new random rotation in every cell it crosses. The largest angular velocity is $\omega_0=D_r\Tx/k_BT$. For consistency with the force case, the figure legends quote $\Tx/v_0 = \zeta$, although, as we show below, the physically relevant comparison for torques is between $\omega_0$ and the rates $D_r$ and $v_0/a$.

We study $\Fx=0$, $125$, $160$, $178$, $187$, $200$, and $250$ ($\chi$ from $0$ to $2$), and $\Tx=0$, $10$ $15$, $20$, $30$, and $100$ ($\zeta$ from $0$ to $0.8$).

\subsection{Simulation protocol}

Equations~\eqref{eq:eom_r} and \eqref{eq:eom_theta} are integrated with the Euler-Maruyama scheme and time step $\Delta t=10^{-5}$, using an in-house OpenMP-parallel Fortran code at fixed $N$ and $L$. Initial configurations are prepared at a high temperature $k_BT=6$ and weak activity $v_0=10$, where the system is homogeneous. At $t=0$ we switch to $k_BT=1$, $v_0=125$, and the chosen disorder. All quantities are averaged over 30 independent initial configurations.
% AUTHORS: state whether each run uses a new disorder realization.

\subsection{Observables}
\label{sec:observables}

\emph{Single-particle dynamics.} We measure the mean-squared displacement (MSD) and the mean-squared angular displacement (MSAD),
\begin{align}
\mathrm{MSD}(t)&=\left\langle|\mathbf r_i(t_0+t)-\mathbf r_i(t_0)|^2\right\rangle,\\
\mathrm{MSAD}(t)&=\left\langle[\theta_i(t_0+t)-\theta_i(t_0)]^2\right\rangle,
\end{align}
where $\theta_i$ is unwrapped and the average runs over particles and runs. For a free ABP, $\mathrm{MSAD}=2D_rt$ and the MSD crosses over from ballistic ($\propto v_0^2t^2$) to diffusive behavior on the persistence time $1/D_r$.

\emph{Order parameter.} To characterize the morphology we coarse-grain the system into square cells of area $(2\sigma)^2$ and assign to each cell the order parameter
\begin{equation}
\psi(\mathbf r,t)=
\begin{cases}
+1, & \text{local area fraction}>\phi_c,\\
-1, & \text{otherwise},
\end{cases}
\label{eq:psi}
\end{equation}
which distinguishes particle-rich from particle-poor regions.
% AUTHORS: give the value of the threshold phi_c.

\emph{Correlation function and structure factor.} The equal-time correlation function is \cite{Bray1994,ahmad2010kinetics}
\begin{equation}
C(r,t)=\langle\psi(\mathbf 0,t)\psi(\mathbf r,t)\rangle-\langle\psi(\mathbf 0,t)\rangle\langle\psi(\mathbf r,t)\rangle,
\label{eq:C}
\end{equation}
averaged over origins, directions, and runs, and the structure factor is its Fourier transform \cite{bbhy-gmsp},
\begin{equation}
S(k,t)=\langle|\psi_{\mathbf k}(t)|^2\rangle=\int C(\mathbf r,t)\,e^{i\mathbf k\cdot\mathbf r}\,d^2r,
\label{eq:S}
\end{equation}
averaged over the directions of $\mathbf k$. If the morphology is statistically self-similar, with a single time-dependent length $\ell(t)$, both quantities obey dynamical scaling~\cite{Bray1994,Puri2009},
\begin{equation}
C(r,t)=\widetilde C\!\left(\frac{r}{\ell(t)}\right),\qquad S(k,t)=\ell^d\,\widetilde S\big(k\ell(t)\big),
\label{eq:scaling}
\end{equation}
with $d=2$. We define $\ell(t)$ as the first distance at which $C(r,t)$ falls to $0.1$. Other thresholds change $\ell$ only by a constant factor and do not affect exponents.

For a conserved order parameter with sharp interfaces, the scaling functions have well-known features~\cite{Bray1994,Yeung1988}. (a) The sum rule $\int C(\mathbf r,t)\,d^2r=0$, which follows from conservation, forces $C$ to become negative, producing a pronounced minimum. (b) Sharp interfaces give a linear short-distance decay, $C\simeq1-a_0r/\ell$ (Porod behavior), and correspondingly a Porod tail $S(k)\sim k^{-(d+1)}=k^{-3}$ at large $k\ell$. (c) At small $k$, $S\sim k^4$. Departures from (a) and (b) are the main morphological signatures of disorder that we look for below.

\emph{Effective exponents.} To characterize the growth kinetics without assuming a power law, we compute the effective growth exponent and effective dynamic exponent,
\begin{equation}
\alpha_{\mathrm{eff}}=\frac{d\ln\ell}{d\ln t},\qquad \zeff=\frac{1}{\alpha_{\mathrm{eff}}}=\frac{d\ln t}{d\ln\ell},
\label{eq:zeff_def}
\end{equation}
from smoothed $\ell(t)$ data, and plot $\zeff$ against $\ell$. For a power law $\ell\sim t^{1/z}$, $\zeff=z$ is constant, and LS growth corresponds to $z=3$.

%===========================================================
\section{Random force field}
\label{sec:force}
%===========================================================

\subsection{Real-space morphology}

\begin{figure}[t]
\centering
\includegraphics[width=\columnwidth]{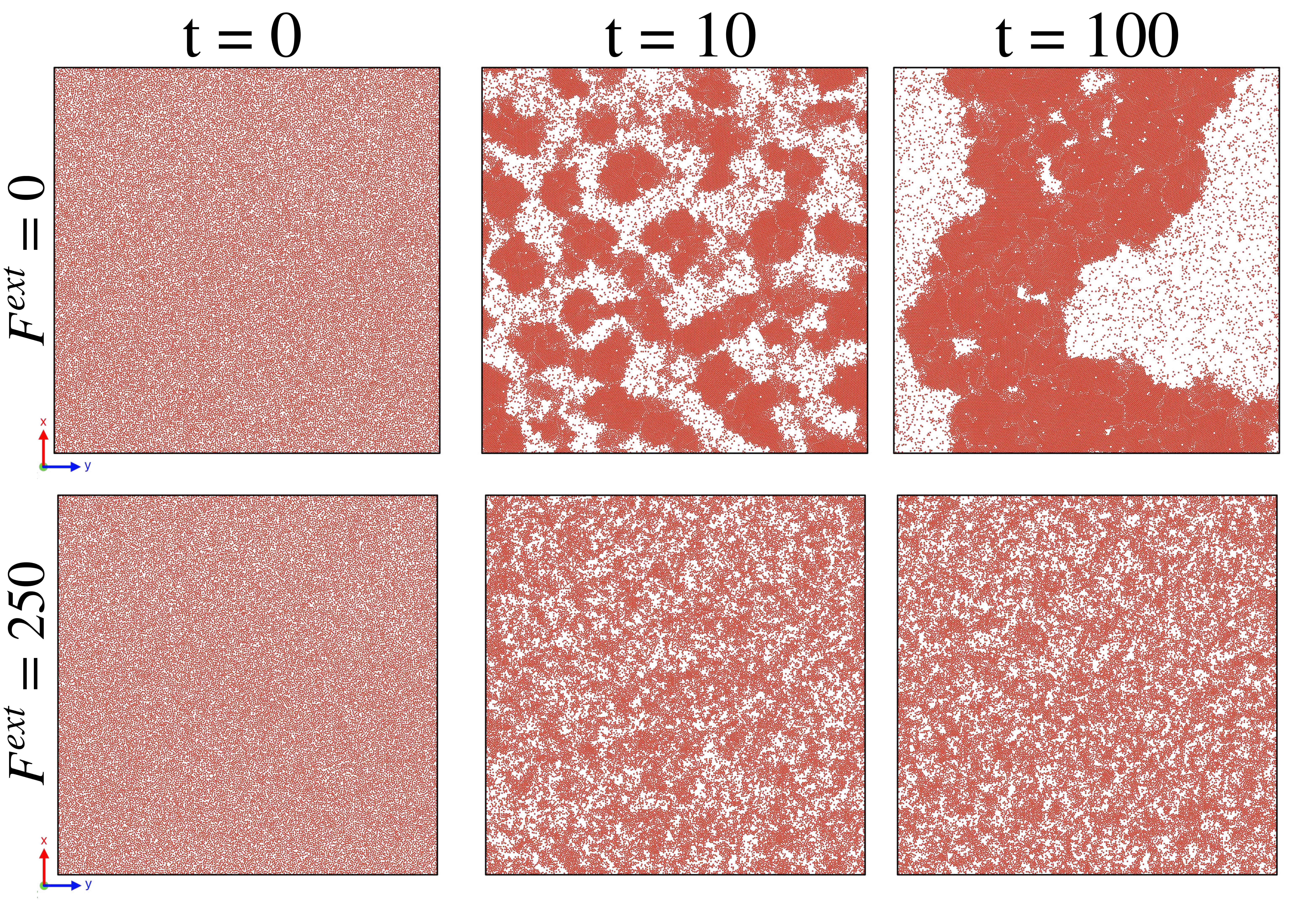}
\caption{Snapshots of the system at $t=0$, $10$, and $100$ for the clean system, $\Fx=0$ (top), and with a random force field, $\Fx=250$ (bottom). Without disorder, dense clusters form and merge into a single slab-like domain. With disorder, the pattern freezes at a scale of a few particle diameters.}
\label{fig:time_evolution}
\end{figure}

Figure~\ref{fig:time_evolution} compares the evolution of the clean system with that of a system in a strong random force field, $\Fx=250$ ($\chi=2$). In the clean case, the homogeneous initial state becomes unstable within a short time. Dense clusters are clearly visible at $t=10$, and by $t=100$ they have merged into a single slab-like domain spanning the box, separated from the gas by fairly sharp, smooth interfaces. This is the familiar MIPS scenario.

With disorder the picture is very different. The system still demixes locally, and dense and dilute patches appear by $t=10$. But these patches remain small and irregular, and the pattern at $t=100$ is hardly distinguishable from that at $t=10$. No compact domains form and no sharp interfaces develop.

At weaker forcing the effect is gradual. The disorder first roughens the cluster boundaries while leaving the phase-separated morphology intact. As $\Fx$ approaches and exceeds $v_0$, clusters fragment and the pattern becomes increasingly fine and irregular. We quantify these observations below, starting from the motion of individual particles.

\subsection{Single-particle dynamics}

\begin{figure}[t]
\centering
\includegraphics[width=0.45\textwidth,height=0.4\textwidth]{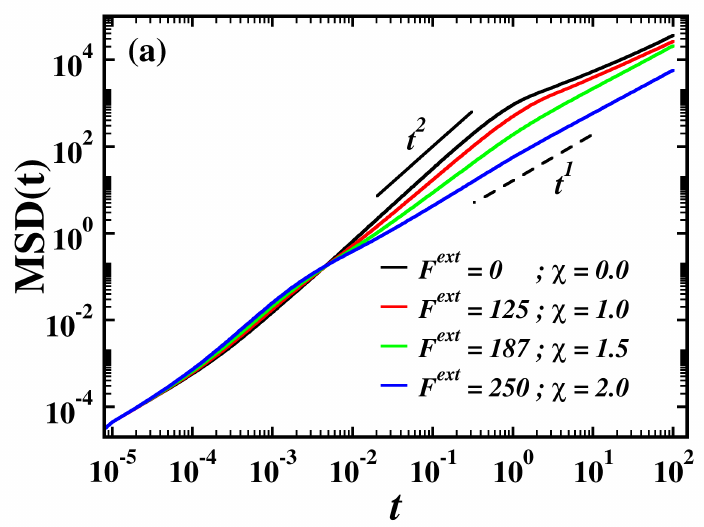}
\includegraphics[width=0.45\textwidth,height=0.4\textwidth]{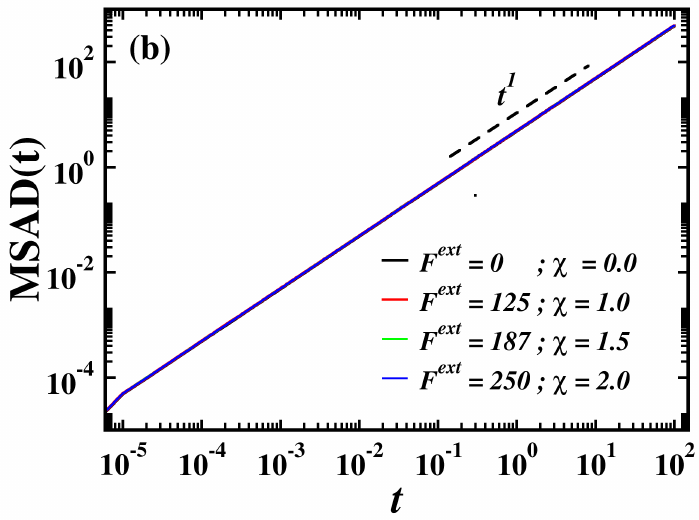}
\caption{Single-particle dynamics in a random force field. (a) Mean-squared displacement for $\Fx/v_0=\chi=0$ to $2$. The ballistic regime ($\propto t^2$, solid line) ends earlier and the long-time diffusive regime ($\propto t$, dashed line) is slower as $\Fx$ increases. (b) Mean-squared angular displacement for the same forces. All curves coincide and grow as $2D_rt$.}
\label{fig:MSD_MSAD}
\end{figure}

Figure~\ref{fig:MSD_MSAD} shows the MSD and MSAD. The MSAD [Fig.~\ref{fig:MSD_MSAD}(b)] is the same for all force strengths and grows linearly, $\mathrm{MSAD}=2D_rt$, at all times. This is expected, since $\mathbf f$ enters only the translational equation~\eqref{eq:eom_r}. Random forces leave the orientational dynamics, and hence the intrinsic persistence time $1/D_r$, completely unchanged.

The translational dynamics, by contrast, depend strongly on $\Fx$ [Fig.~\ref{fig:MSD_MSAD}(a)]. At very short times all curves coincide, since the dynamics are dominated by thermal noise. On intermediate time scales the MSD is ballistic, $\mathrm{MSD}\sim t^2$, reflecting persistent self-propulsion, and at long times it becomes diffusive, $\mathrm{MSD}\sim t$. As $\Fx$ grows, the ballistic regime becomes shorter, the crossover to diffusion occurs earlier, and the long-time diffusivity decreases. The effective mobility of particles is thus strongly reduced.

The natural measure of the force disorder is the ratio of the local drift speed $|\mathbf f|$ to the swim speed $v_0$. As shown in Sec.~\ref{sec:mech_force}, for $\Fx\ge v_0$ a fraction $p=1-\pi v_0^2/[4(\Fx)^2]$ of all cells imposes a drift larger than $v_0$ (Table~\ref{tab:drift}). In such cells a particle cannot swim against the local flow, whatever its orientation. For $\Fx=250$, this is the case in $80\%$ of the cells. Since MIPS relies on persistent motion and repeated collisions to sustain accumulation, this loss of mobility directly weakens the self-trapping mechanism.

\begin{table}[b]
\caption{Random forces: disorder strength, its ratio to the swim speed, and the fraction $p$ of lattice cells in which the local drift speed exceeds $v_0$, Eq.~\eqref{eq:p}.}
\label{tab:drift}
\begin{ruledtabular}
\begin{tabular}{cccccc}
$\Fx$ & 125 & 178 & 187 & 200 & 250\\
$\Fx/v_0$ & 1.00 & 1.42 & 1.50 & 1.60 & 2.00\\
$p$ & 0.21 & 0.61 & 0.65 & 0.69 & 0.80\\
\end{tabular}
\end{ruledtabular}
\end{table}

\subsection{Correlation function and the breakdown of superuniversality}

\begin{figure*}[t]
\centering
\includegraphics[width=0.45\textwidth,height=0.4\textwidth]{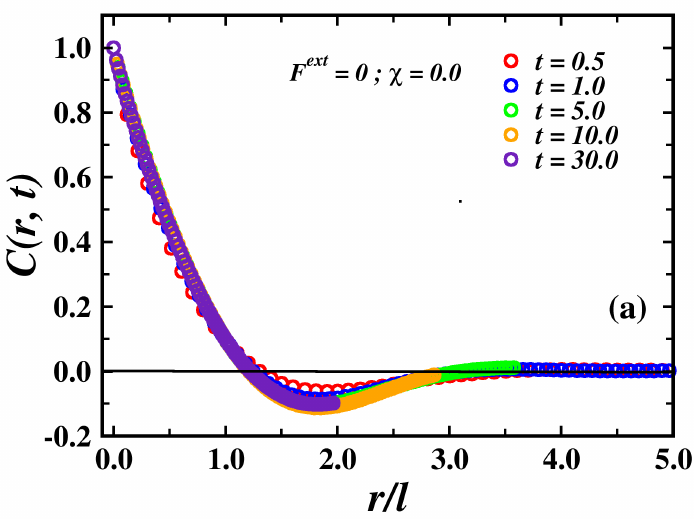}
\includegraphics[width=0.45\textwidth,height=0.4\textwidth]{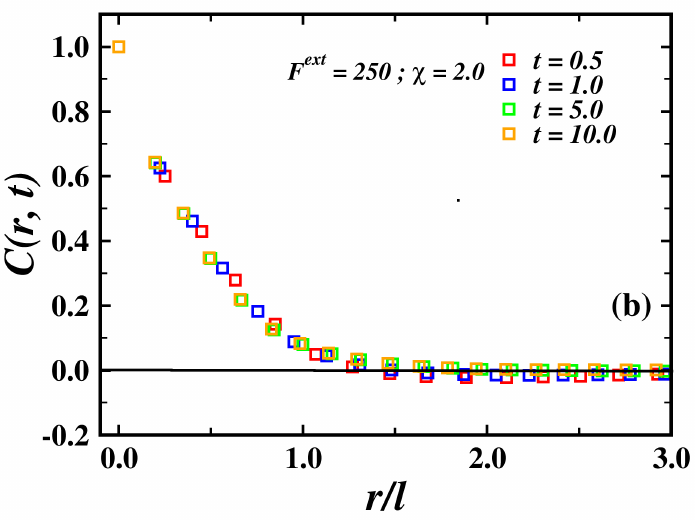}
\includegraphics[width=0.45\textwidth,height=0.4\textwidth]{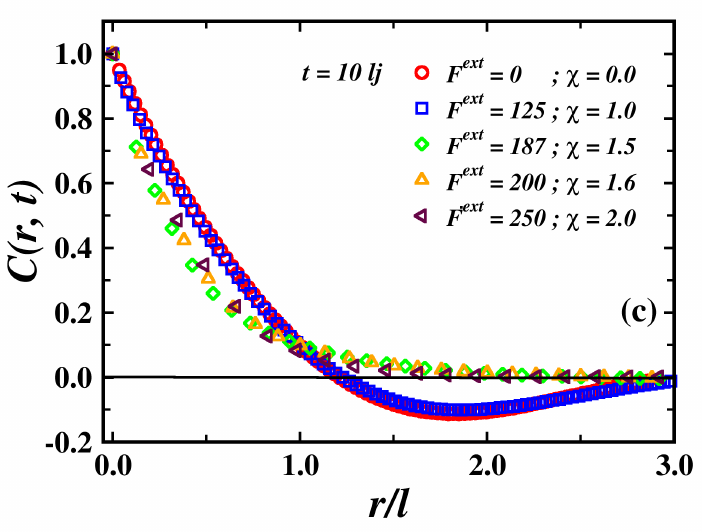}
\caption{Equal-time correlation function for random forces. (a),(b) $C(r,t)$ versus $r/\ell(t)$ at several times for $\Fx=0$ and $\Fx=250$. In both cases the data collapse, showing dynamical scaling. (c) Scaled correlation functions for $\Fx=0$ to $250$. The curves do not collapse onto a single master curve: superuniversality fails. With increasing disorder the negative minimum disappears and the short-distance decay is no longer linear.}
\label{fig:force_cor}
\end{figure*}

Figures~\ref{fig:force_cor}(a) and \ref{fig:force_cor}(b) show $C(r,t)$ as a function of $r/\ell(t)$ at several times for $\Fx=0$ and $\Fx=250$. In both cases the data for different times collapse onto a single curve. Coarsening is therefore self-similar, and $\ell(t)$ is the only relevant time-dependent length, even in strong disorder. We have checked that the same holds at all force strengths studied.

The shape of the master curve, however, changes markedly with disorder. In the clean system, $C$ drops linearly at small $r/\ell$, crosses zero, and develops a deep negative minimum near $r\approx2\ell$ before decaying to zero with weak oscillations. These are the textbook features of compact domains with sharp interfaces and a conserved order parameter. For $\Fx=250$ the minimum is almost gone, the decay at small $r$ is curved, and $C$ approaches zero monotonically.

Figure~\ref{fig:force_cor}(c) compares the scaling functions for all force strengths. They do not collapse. With increasing $\Fx$ the profiles become broader at intermediate distances, the negative dip becomes shallower and eventually vanishes, and the initial decay loses its linear form. The scaling function thus depends explicitly on the disorder strength, which means that superuniversality~\cite{Fisher1988,Paul2004} does not hold.

The disappearance of the negative dip deserves comment. The number of particles is still conserved, and so is the global order parameter; the sum rule $\int C\,d^2r=0$ therefore still holds. A correlation function without a visible negative region means that the compensating negative weight is spread out to large distances, rather than concentrated in a depletion zone around each domain. This is what one expects if dense regions are anchored to favorable locations of the disorder instead of being fed by diffusion from their immediate neighborhood. It is also consistent with the long-range, disorder-induced correlations predicted for scalar active matter in quenched random environments~\cite{Ro2021}. At strong forcing the morphology thus superficially resembles that of a nonconserved system, although the dynamics remain conserved.

\subsection{Structure factor}

\begin{figure}[t]
\centering
\includegraphics[width=0.45\textwidth,height=0.4\textwidth]{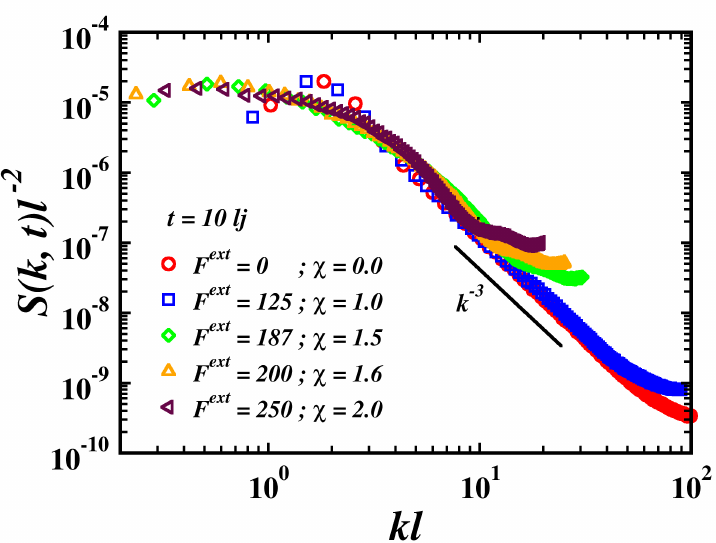}
\caption{Scaled structure factor $S(k,t)\ell^{-2}$ versus $k\ell$ on log-log scales for $\Fx=0$ to $250$. The solid line is the Porod law $S\sim k^{-3}$. With increasing disorder the peak flattens and the large-$k$ tail becomes less steep than $k^{-3}$.}
\label{fig:force_struc}
\end{figure}

The structure factor (Fig.~\ref{fig:force_struc}) tells the same story in reciprocal space. In the clean system, $S(k)\ell^{-2}$ has a peak at $k\ell$ of order one, which reflects the dominant domain size, and a large-$k$ tail consistent with the Porod law $S\sim k^{-3}$. With increasing $\Fx$, the tail becomes progressively less steep, which signals rough and diffuse interfaces. The peak also flattens and eventually disappears, so the morphology no longer has a well-defined characteristic wavelength. At the largest forces the high-$k$ tail develops a shoulder, reflecting density fluctuations on the scale of a few particle diameters imprinted by the disorder.

One caveat applies. At the strongest disorder, $\ell$ is only a few coarse-graining cells, so the largest $k\ell$ values are partly limited by the coarse-graining. The departure from the Porod law is, however, already clear at intermediate forces, where $\ell$ is larger.

\subsection{Growth kinetics}

\begin{figure}[t]
\centering
\includegraphics[width=0.45\textwidth,height=0.4\textwidth]{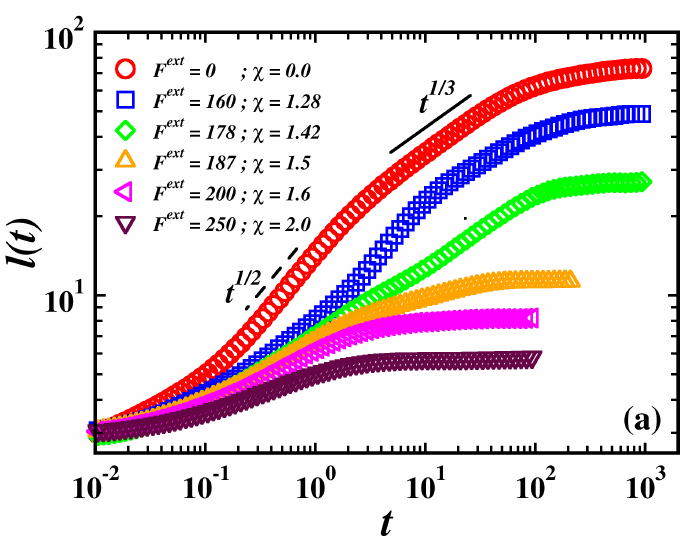}

\includegraphics[width=0.45\textwidth,height=0.4\textwidth]{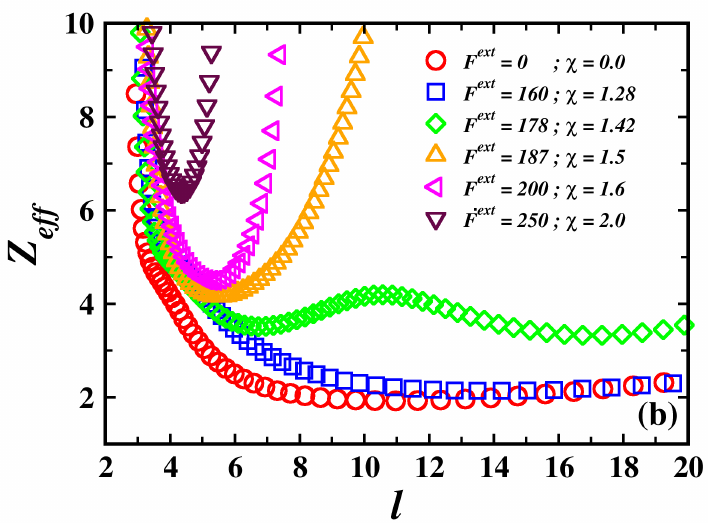}
\caption{Growth kinetics in a random force field. (a) Domain size $\ell(t)$ on log-log scales. The legend gives $\Fx$ and $\Fx/v_0=\chi$. The clean system follows the Lifshitz-Slyozov law $\ell\sim t^{1/3}$; with disorder, growth is slower and the curves bend over at late times. (b) Effective dynamic exponent $\zeff=d\ln t/d\ln\ell$ versus $\ell$ for $\Fx=0$ to $250$.}
\label{fig:force_length}
\end{figure}

Figure~\ref{fig:force_length}(a) shows the domain size as a function of time. For $\Fx=0$ the growth follows $\ell\sim t^{\alpha}$ with $\alpha\simeq1/3$ over the whole run, as expected for LS coarsening. With disorder, growth is slower at all times, and the curves bend over at late times. For $\Fx=250$, $\ell$ saturates at about $5.5\sigma$ after $t\approx1$. For intermediate forces the curves are close to power laws over a limited range, before they flatten.

A log-log plot alone cannot distinguish a smaller growth exponent, a slower-than-algebraic growth, and complete arrest. We therefore turn to the effective dynamic exponent, Eq.~\eqref{eq:zeff_def}, shown as a function of $\ell$ in Fig.~\ref{fig:force_length}(b). For a power law $\zeff$ would be flat. Instead, each curve decreases to a minimum $\bar z$ and then turns upward steeply. For the weakest disorder shown, $\Fx=178$, the minimum is shallow and wide, with $\bar z\approx2.9$ close to the LS value, and the upturn only sets in beyond $\ell\approx30$. As $\Fx$ increases, the minimum moves up, becomes narrower, and the upturn moves to smaller $\ell$.

The intermediate regime is therefore a power law with a disorder-dependent exponent $1/\bar z$, much as found in simulations of random magnets~\cite{Paul2004}. The late-time upturn shows that this power law is only transient. The growth becomes progressively slower as domains grow, which is the signature of activated dynamics.

\subsection{Crossover scaling and logarithmic growth}

\begin{figure*}[t]
\centering
\includegraphics[width=0.45\textwidth,height=0.4\textwidth]{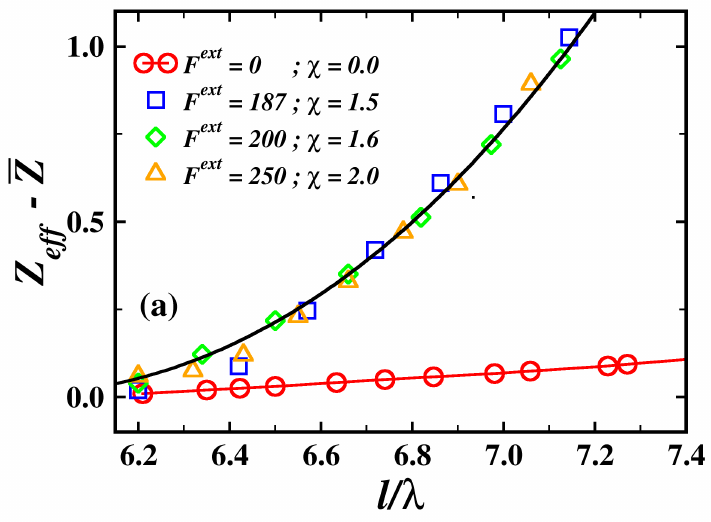}\hfill
\includegraphics[width=0.45\textwidth,height=0.4\textwidth]{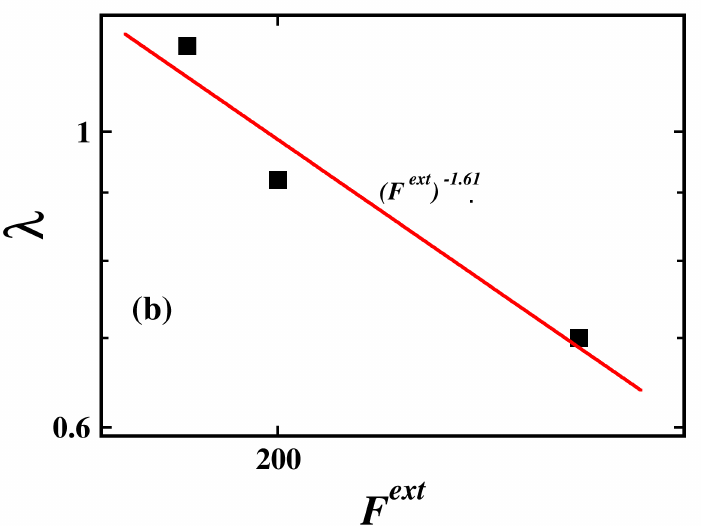}\hfill
\includegraphics[width=0.45\textwidth,height=0.4\textwidth]{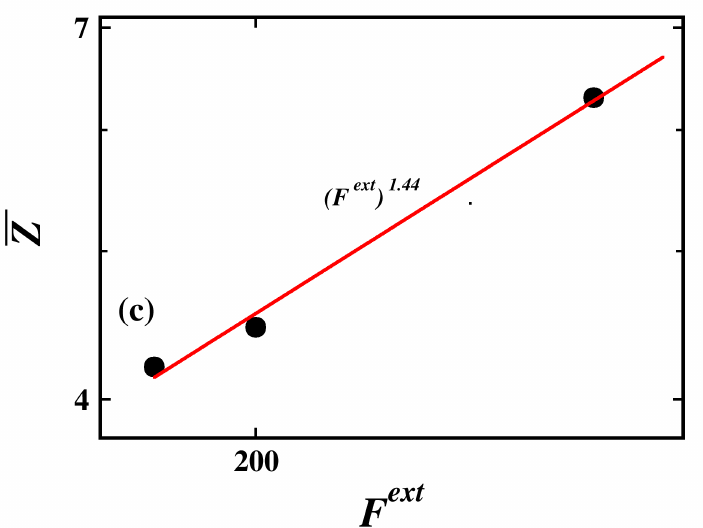}
\caption{Crossover scaling for random forces. (a) Collapse of $\zeff-\bar z$ against $\ell/\lambda$ for $\Fx=187$, $200$, and $250$; the solid black line is Eq.~\eqref{eq:zeff_scaling}. (b) Crossover length $\lambda$ versus $\Fx$ on log-log scales; the solid red line has slope $-1.61$. (c) Plateau exponent $\bar z$ versus $\Fx$; the solid red line has slope $1.44$.}
\label{fig:force_cross}
\end{figure*}

To describe the crossover quantitatively, we follow Corberi \emph{et al.}, who proposed for the random-field Ising model the form~\cite{Corberi2012}
\begin{equation}
\zeff=\bar z+b\left(\frac{\ell-\ell_0}{\lambda}\right)^{\varphi},
\label{eq:zeff_scaling}
\end{equation}
where $\bar z$ is the plateau value, $\lambda$ a disorder-dependent crossover length, and $b$, $\ell_0$, and $\varphi$ are constants. We choose $\lambda$ for each disorder strength so as to collapse $\zeff-\bar z$ as a function of $\ell/\lambda$. The result, shown in Fig.~\ref{fig:force_cross}(a), is a good collapse for $\Fx=187$, $200$, and $250$, and the collapsed data are well described by Eq.~\eqref{eq:zeff_scaling} with $b\simeq0.463$, $\ell_0\simeq5.85$, and $\varphi\simeq2.238$ (Table~\ref{tab:fit}). The slowing down is thus controlled by one disorder-dependent length scale, even though the system is far from equilibrium.
% AUTHORS: the collapse axis is l/lambda (13-15), so please check that b and l_0 correspond to Eq. (zeff_scaling) as written.

The crossover length decreases with increasing disorder [Fig.~\ref{fig:force_cross}(b)], approximately as $\lambda\sim(\Fx)^{-1.61}$, while the plateau exponent increases, $\bar z\sim(\Fx)^{1.44}$ [Fig.~\ref{fig:force_cross}(c)]. Stronger disorder therefore makes the transient growth slower and brings the activated regime to smaller domain sizes.

Equation~\eqref{eq:zeff_scaling} has a transparent physical meaning, derived in Appendix~\ref{app:growth}. Integrating $d\ln t=\zeff\,d\ln\ell$ gives, for $\ell\gg\ell_0$,
\begin{equation}
t(\ell)\simeq t_0\,\left(\frac{\ell}{\ell_0}\right)^{\bar z}\exp\!\left[\frac{b}{\varphi}\left(\frac{\ell}{\lambda}\right)^{\varphi}\right].
\label{eq:arrhenius}
\end{equation}
The first factor is the transient power law. The second factor has an Arrhenius-like form, $\exp[E(\ell)/T_{\mathrm{eff}}]$, with a barrier that grows with domain size as
\begin{equation}
\frac{E(\ell)}{T_{\mathrm{eff}}}=\frac{b}{\varphi}\left(\frac{\ell}{\lambda}\right)^{\varphi}.
\label{eq:barrier}
\end{equation}
This form is analogous to the activated domain-growth form proposed by Huse and Henley for domain growth in random magnets, where pinned domain walls encounter barriers that increase with domain size~\cite{Huse1985}. Inverting Eq.~\eqref{eq:arrhenius} at late times gives the growth law
\begin{equation}
\ell(t)\simeq\lambda\left(\frac{\varphi}{b}\ln t\right)^{1/\varphi},
\label{eq:log_growth}
\end{equation}
so that growth becomes logarithmic once $\ell$ is well beyond the crossover (Appendix~\ref{app:growth}). With $\varphi\approx2.2$, this growth is extremely slow. Over any practical time window it cannot be distinguished from arrest, and in two dimensions the Imry-Ma argument of Ro \emph{et al.}~\cite{Ro2021} suggests that phase separation is indeed limited to a finite scale. Whether $\ell(t)$ ultimately saturates or keeps creeping logarithmically, the mechanism is the same and is qualitatively different from LS coarsening: growth proceeds through rare, activated rearrangements of pinned interfaces rather than through diffusive evaporation and condensation.

%===========================================================
\section{Random torque field}
\label{sec:torque}
%===========================================================

We now turn to orientational disorder, with $\Fx=0$ and $\Tx>0$. In contrast with random forces, random torques do not act on positions at all. They act on the orientation, and thus directly on the persistence of self-propelled motion, which is the key ingredient of MIPS.

\subsection{Real-space morphology}

\begin{figure}[t]
\centering
\includegraphics[width=\columnwidth]{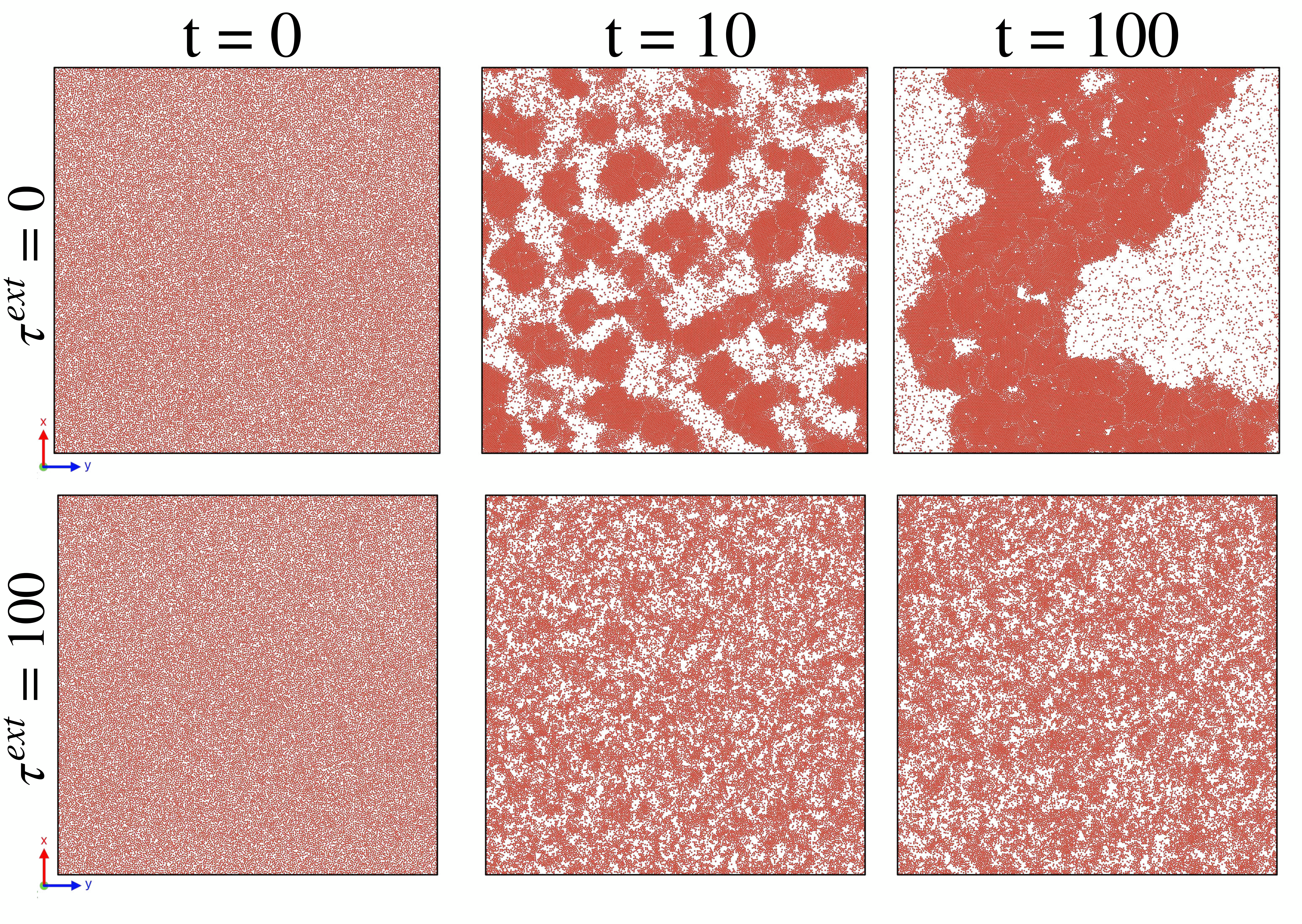}
\caption{Snapshots at $t=0$, $10$, and $100$ for the clean system, $\Tx=0$ (top), and with a random torque field, $\Tx=100$ (bottom). With random torques the system stays close to homogeneous, with only small, short-lived density fluctuations.}
\label{fig:time_evolution_torq}
\end{figure}

Figure~\ref{fig:time_evolution_torq} compares the clean system with a system in a random torque field of strength $\Tx=100$. The contrast is even sharper than for random forces. The torque-disordered system forms only small, weak density fluctuations that do not grow, and it looks nearly homogeneous at all times. Note that the torque amplitude here, $\Tx/v_0=0.8$, is less than half the force amplitude used in Fig.~\ref{fig:time_evolution}. Increasing $\Tx$ from zero, we observe a gradual change from the clustered to a dispersed state.

\subsection{Single-particle dynamics}

\begin{figure}[t]
\centering
\includegraphics[width=0.45\textwidth,height=0.4\textwidth]{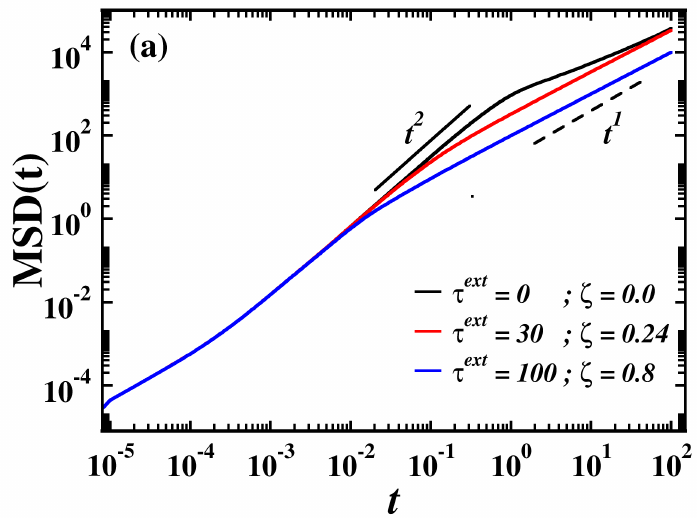}
\includegraphics[width=0.45\textwidth,height=0.4\textwidth]{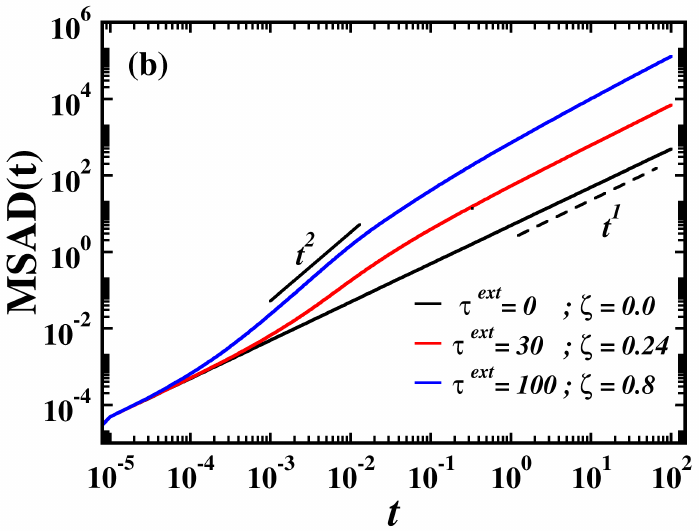}
\caption{Single-particle dynamics in a random torque field for $\Tx=0$, $30$, and $100$ ($\Tx/v_0=\zeta=0$, $0.24$, $0.8$). (a) Mean-squared displacement: the MSD at intermediate and long times decreases with increasing $\Tx$. (b) Mean-squared angular displacement: rotation becomes ballistic, $\propto t^2$ (solid line), at short times and much faster than $2D_rt$ at long times. Dashed lines have slope $1$.}
\label{fig:torq_MSD_MSAD}
\end{figure}

The single-particle dynamics are shown in Fig.~\ref{fig:torq_MSD_MSAD}. The MSD [Fig.~\ref{fig:torq_MSD_MSAD}(a)] shows the usual crossover from ballistic to diffusive motion, but its magnitude at intermediate and long times drops markedly with increasing $\Tx$. Particles cover less ground because they lose their direction of motion more quickly.

The MSAD [Fig.~\ref{fig:torq_MSD_MSAD}(b)] shows this loss of orientation directly, and in a revealing way. For $\Tx=0$ it grows as $2D_rt$. For $\Tx=100$ it instead grows as $t^2$ at short times, and it is much larger than $2D_rt$ at all times. Rotational motion becomes ballistic before it becomes diffusive. This is the signature of particles rotating at a fixed angular velocity, namely the local $\omega=D_r\Gamma/k_BT$ of the cell they occupy. As shown in Sec.~\ref{sec:mech_torque} and Appendix~\ref{app:dreff}, a parameter-free estimate,
\begin{equation}
\mathrm{MSAD}(t)\simeq2D_rt+\frac{\omega_0^2}{3}\,t^2,\qquad \omega_0=\frac{D_r\Tx}{k_BT},
\label{eq:msad_short}
\end{equation}
reproduces the short-time behavior. At $t\approx10^{-2}$ it predicts $\mathrm{MSAD}\approx2$ for $\Tx=100$ and $\approx0.23$ for $\Tx=30$, in agreement with Fig.~\ref{fig:torq_MSD_MSAD}(b). At long times the rotational dynamics become diffusive again, but with a strongly enhanced effective rotational diffusion coefficient.

The combination of shorter persistence and reduced translational mobility has direct consequences for MIPS, since self-trapping requires particles to keep pushing in the same direction long enough for others to pile up behind them. Random torques disrupt exactly this.

\subsection{Correlation function}

\begin{figure}[t]
\centering
\includegraphics[width=0.45\textwidth,height=0.4\textwidth]{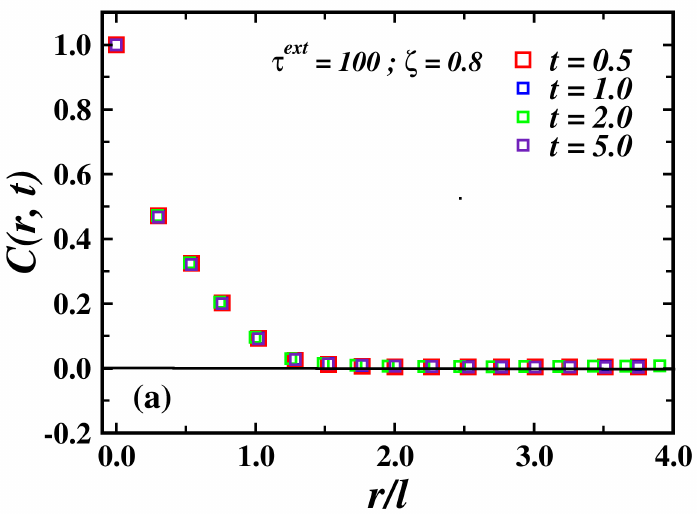}
\includegraphics[width=0.45\textwidth,height=0.4\textwidth]{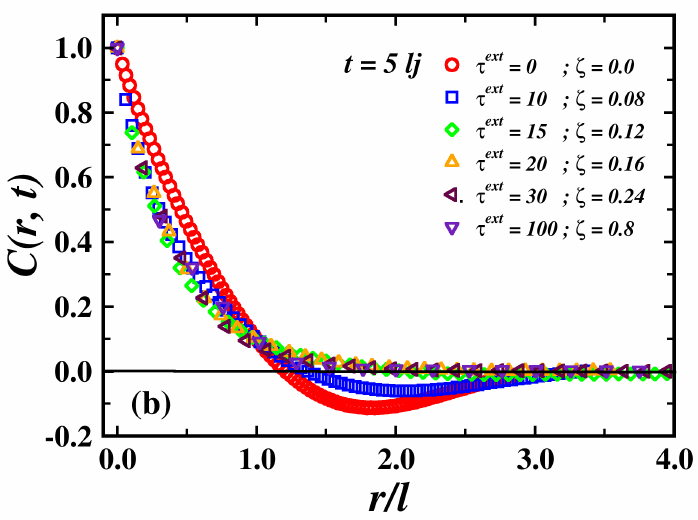}
\caption{Equal-time correlation function for random torques. (a) $C(r,t)$ versus $r/\ell(t)$ at several times for strong torque disorder; the data collapse. (b) Scaled correlation functions for $\Tx=0$ to $100$. The curves do not collapse; the negative minimum of the clean system disappears for all $\Tx>10$. The clean case at several times is shown in Fig.~\ref{fig:force_cor}(a).}
\label{fig:torque_cor}
\end{figure}
% AUTHORS: file torq_30_cor suggests tau_ext = 30, while the original caption said tau_ext = 100; the caption above avoids a value until this is checked.

For fixed torque strength, $C(r,t)$ again collapses when plotted against $r/\ell(t)$ [Fig.~\ref{fig:torque_cor}(a)], so coarsening remains self-similar. The scaling functions for different $\Tx$ are compared in Fig.~\ref{fig:torque_cor}(b). As for random forces, they do not collapse, and superuniversality fails. The changes, however, are more abrupt. Already at the smallest torque studied, $\Tx=15$ ($\Tx/v_0=\zeta=0.12$), the negative dip has disappeared and $C$ decays faster at short distances than in the clean system. Further increases of $\Tx$ change the curve only moderately. The short-distance decay is no longer linear, which again indicates rough and diffuse interfaces.

Random torques therefore produce the same qualitative morphological signatures as random forces: loss of the conserved-order-parameter dip, breakdown of Porod behavior, and disorder-dependent scaling functions. But they do so at much smaller amplitudes.

\subsection{Structure factor}

\begin{figure}[t]
\centering
\includegraphics[width=0.45\textwidth,height=0.4\textwidth]{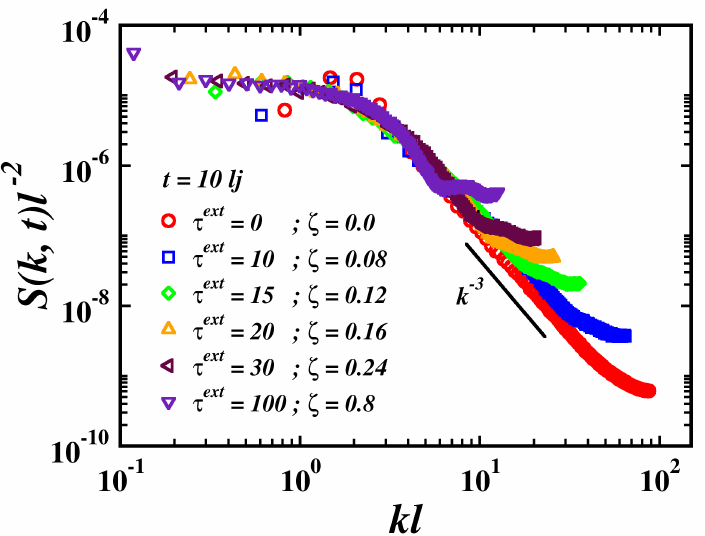}
\caption{Scaled structure factor $S(k,t)\ell^{-2}$ versus $k\ell$ for $\Tx=0$ to $100$. The solid line is the Porod law $k^{-3}$.}
\label{fig:torque_struc}
\end{figure}

The scaled structure factors (Fig.~\ref{fig:torque_struc}) confirm this. For $\Tx=0$ the Porod law holds at large $k\ell$. For all $\Tx>0$ the tail becomes less steep and develops a shoulder at large $k\ell$, the structure-factor counterpart of rough interfaces and fine-scale density fluctuations. The peak at $k\ell\sim1$ is weakened. The same caveat about coarse-graining at large $k\ell$ applies as for random forces.

\subsection{Growth kinetics}

\begin{figure}[t]
\centering
%\panel{0.85\columnwidth}{Figure_5.png}{a}\\[1mm]
%\panel{0.85\columnwidth}{torq_1.png}{b}
\includegraphics[width=0.45\textwidth,height=0.4\textwidth]{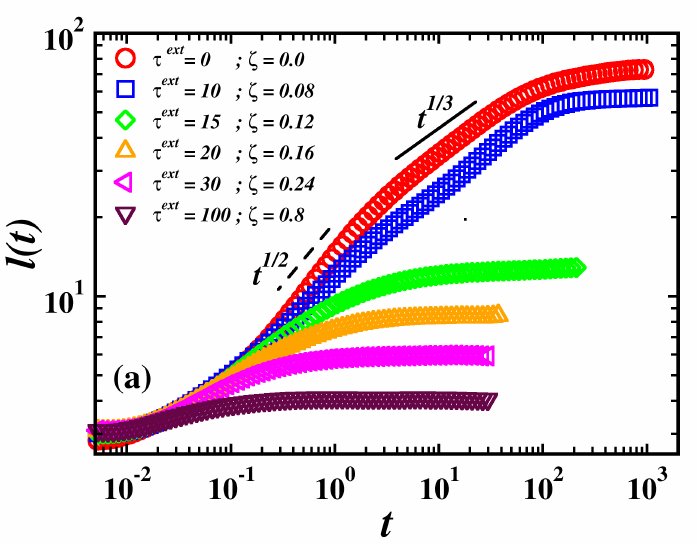}
\includegraphics[width=0.45\textwidth,height=0.4\textwidth]{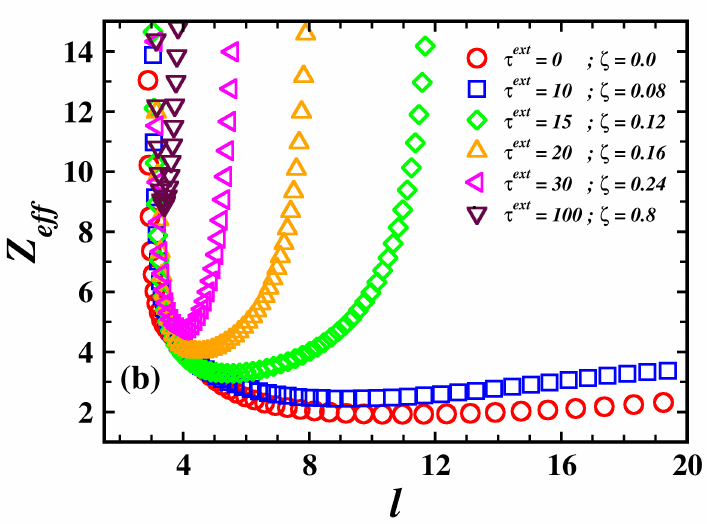}
\caption{Growth kinetics in a random torque field. (a) Domain size $\ell(t)$ on log-log scales; the legend gives $\Tx$ and $\Tx/v_0$ (``$\zeta$''). (b) Effective dynamic exponent $\zeff$ versus $\ell$ for $\Tx=0$, $10$, $15$, $20$, $30$, and $100$.}
\label{fig:torq_length}
\end{figure}

The domain size is shown in Fig.~\ref{fig:torq_length}(a). The clean system follows LS growth, $\ell\sim t^{1/3}$. With torques, growth is slower at all times and saturates at a torque-dependent level: $\ell$ levels off at about $6$-$9\sigma$ for $\Tx=20$-$30$ and at about $4\sigma$ for $\Tx=100$, where it barely grows at all after $t\approx0.1$. Only for the weakest torque, $\Tx=15$, does $\ell$ keep growing noticeably up to the end of the run.

The effective dynamic exponent [Fig.~\ref{fig:torq_length}(b)] shows the same structure as for random forces. Each curve has a minimum $\bar z$ followed by a steep upturn. For $\Tx=15$ the minimum is broad and lies near $\bar z\approx2.9$, close to the LS value. With increasing torque it moves up, and to $\bar z\approx9$ for $\Tx=100$, while the upturn moves to smaller $\ell$.

\subsection{Crossover scaling}

\begin{figure*}[t]
\centering
%\panel{0.4\textwidth}{torq_2.png}{a}\hspace{0.05\textwidth}
%\panel{0.4\textwidth}{torq_3.png}{b}
\includegraphics[width=0.45\textwidth,height=0.4\textwidth]{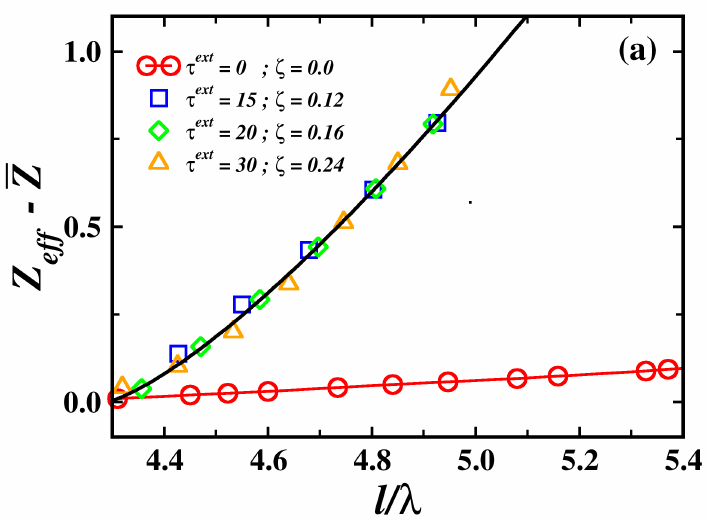}
\includegraphics[width=0.45\textwidth,height=0.4\textwidth]{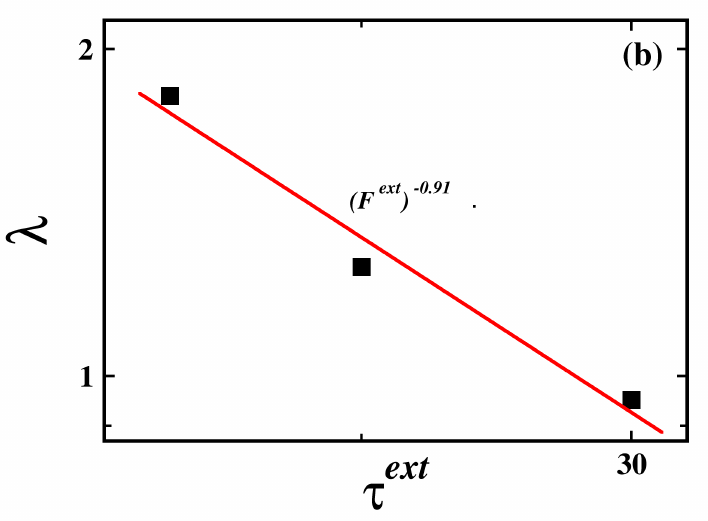}
\includegraphics[width=0.45\textwidth,height=0.4\textwidth]{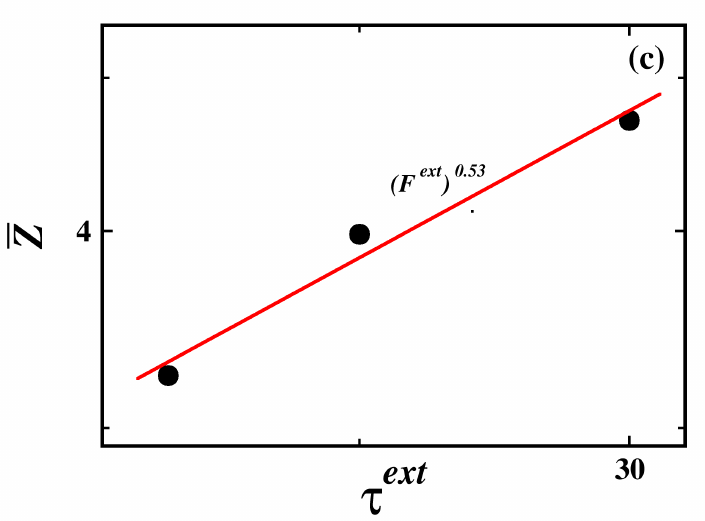}
\caption{Crossover scaling for random torques. (a) Collapse of $\zeff-\bar z$ against $\ell/\lambda$ for $\Tx=15$, $20$, and $30$; the solid black line is Eq.~\eqref{eq:zeff_scaling}. (b) Crossover length $\lambda$ versus $\Tx$ on log-log scales; the solid red line has slope $-0.91$.  (c) Plateau exponent $\bar z$ versus $\Tx$; the solid red line has slope $0.53$.}
\label{fig:torq_cross}
\end{figure*}

Applying the same scaling analysis, we obtain a good collapse of $\zeff-\bar z$ against $\ell/\lambda$ for $\Tx=15$, $20$, and $30$ [Fig.~\ref{fig:torq_cross}(a)], described by Eq.~\eqref{eq:zeff_scaling} with $b\simeq1.97$, $\ell_0\simeq4.28$, and $\varphi\simeq1.33$. The crossover length decreases with torque strength, approximately as $\lambda\sim(\Tx)^{-0.91}$ [Fig.~\ref{fig:torq_cross}(b)]. By Eqs.~\eqref{eq:arrhenius}-\eqref{eq:log_growth}, the late-stage growth is again activated, with barriers growing as $\ell^{\varphi}$ and domain growth at most logarithmic.

The parameters for both kinds of disorder are collected in Table~\ref{tab:fit}. The barrier exponent is somewhat smaller for torques than for forces, and the crossover length depends less steeply on disorder strength. Although with only one fitted value for each kind of disorder we cannot say whether $\varphi$ is universal. Given the narrow ranges, we do not attach much weight to these quantitative differences. The central observation is that both kinds of disorder lead to the same kind of kinetics: a transient power law with a disorder-dependent exponent, followed by activated, logarithmically slow growth described by the same scaling form.

\begin{table}[b]
\caption{Parameters of the crossover scaling form, Eq.~\eqref{eq:zeff_scaling}, and effective exponents for the disorder dependence of $\lambda$ and $\bar z$.}
\label{tab:fit}
\begin{ruledtabular}
\begin{tabular}{lcc}
 & Random forces & Random torques\\
\hline
Values in collapse & $\Fx=187$-$250$ & $\Tx=15$-$30$\\
$b$ & $0.463$ & $1.97$\\
$\ell_0$ & $5.85$ & $4.28$\\
$\varphi$ & $2.238$ & $1.33$\\
$\lambda$ & $\sim(\Fx)^{-1.61}$ & $\sim(\Tx)^{-0.91}$\\
$\bar z$ & $\sim(\Fx)^{1.44}$ & $\sim(\Tx)^{0.53}$\\
\end{tabular}
\end{ruledtabular}
\end{table}

%===========================================================
\section{Physical mechanisms}
\label{sec:mechanism}
%===========================================================

The two kinds of disorder lead to the same kinetic outcome, but they act on different degrees of freedom and differ strongly in efficiency: torques with $\Tx/v_0=\zeta\approx0.2$ already produce the same slowing down as forces with $\Fx/v_0=\chi\approx1.5$. In this section we explain why.

\subsection{Random forces: trapping and a random field for the density}
\label{sec:mech_force}

Since $f_x$ and $f_y$ are independent and uniform in $[-\Fx,\Fx]$, the point $(f_x,f_y)$ is uniformly distributed in a square of side $2\Fx$. The local drift speed $|\mathbf f|$ exceeds the swim speed $v_0$ outside a disk of radius $v_0$. For $v_0\le\Fx$ the disk lies inside the square, and the fraction of cells with $|\mathbf f|>v_0$ is
\begin{equation}
p=1-\frac{\pi v_0^2}{4(\Fx)^2}.
\label{eq:p}
\end{equation}
The values are listed in Table~\ref{tab:drift}. At $\Fx=v_0$ only about a fifth of the cells beat self-propulsion, but in the range where we observe activated coarsening, $\Fx\ge178$, it is more than half. A particle entering such a cell is carried along by the local drift regardless of its orientation. It then ends up in the sinks of the drift field, where the flow converges and the particle cannot swim out.

These sinks are the key to the random-field character of force disorder. In the decomposition~\eqref{eq:helmholtz}, the gradient part $-\nabla V$ defines a random potential whose minima coincide with the sinks. In Fourier space $V_{\mathbf k}=i\,\mathbf k\cdot\mathbf f_{\mathbf k}/k^2$, and for short-range correlated $\mathbf f$ with variance $\Delta\propto(\Fx)^2$,
\begin{equation}
\langle|V_{\mathbf k}|^2\rangle\simeq\frac{\Delta}{k^2}.
\label{eq:Vk}
\end{equation}
The potential is therefore long-range correlated, and it couples linearly to the local density. This is the density analogue of a random field acting on the magnetization in the random-field Ising model~\cite{Imry1975}. Sinks favor the dense phase and sources the dilute one, so the disorder pins the pattern of dense and dilute regions. This explains the imprint of the disorder on the morphology (Figs.~\ref{fig:force_cor}-\ref{fig:force_struc}) and the pinning of interfaces behind the activated kinetics.

A heuristic Imry-Ma argument~\cite{Imry1975}, in the spirit of Ref.~\cite{Ro2021}, shows why such disorder must eventually limit domain growth. From Eq.~\eqref{eq:Vk}, the typical difference in $V$ between two neighboring regions of size $\ell$ scales as $\delta V(\ell)\sim\sqrt\Delta\,\ell^{(2-d)/2}$. Placing a domain of size $\ell$ on the more favorable of the two gains an energy of order $\ell^d\,\delta V\sim\sqrt\Delta\,\ell^{(d+2)/2}$, while the interface costs $\gamma\ell^{d-1}$, with $\gamma$ an effective interfacial tension. For $d<4$ the disorder term wins at large $\ell$, in agreement with the lower critical dimension $d_c=4$ found in Ref.~\cite{Ro2021}. In $d=2$ it wins beyond
\begin{equation}
\ell_{\mathrm{IM}}\sim\frac{\gamma}{\sqrt\Delta}\propto\frac{1}{\Fx}.
\label{eq:lim}
\end{equation}
Active systems have no free energy, so this argument is only a guide. It does explain why the crossover length $\lambda$ decreases with $\Fx$. The stronger dependence we measure, $\lambda\sim(\Fx)^{-1.61}$, refers to strong forces, $\Fx>v_0$, where the weak-disorder estimate does not apply. It also suggests that forces matter strongly only once they compete with self-propulsion, as Table~\ref{tab:drift} shows.

We note in passing that diffusion of a single particle in a short-range correlated random force field is itself marginal in two dimensions~\cite{Bouchaud1990}. In our system this effect is masked by self-propulsion and interactions, and the long-time MSD remains diffusive (Fig.~\ref{fig:MSD_MSAD}).

\subsection{Random torques: quenched chirality and a reduced P\'eclet number}
\label{sec:mech_torque}

With $\Gamma$ uniform in $[-\Tx,\Tx]$, the local angular velocity has $\langle\omega\rangle=0$ and $\langle\omega^2\rangle=\omega_0^2/3$. Random torques act in two ways, depending on whether a particle is stuck or moving.

\emph{Stuck particles: quenched chirality.} A particle that stays in one cell rotates at the fixed rate $\omega$ of that cell, so that its orientation changes by $\Delta\theta=\omega t+\sqrt{2D_r}\,W(t)$, with $W$ a Wiener process. Averaging over cells gives Eq.~\eqref{eq:msad_short}, with a ballistic term that dominates for $t>6D_r/\omega_0^2$. This is what Fig.~\ref{fig:torq_MSD_MSAD}(b) shows. In a dense cluster, where particles barely move, each particle behaves as a chiral swimmer with its own, quenched rotation rate. Chirality is known to suppress MIPS, because rotating particles escape from clusters more easily~\cite{Liebchen2017,Bickmann2022,Ma2022}. Here the chirality is random in sign and position.

\emph{Moving particles: enhanced rotational diffusion.} A particle moving at speed $v$ spends a time $a/v$ in each cell and picks up a rotation $\omega_ja/v$, with $\omega_j$ independent from cell to cell. As shown in Appendix~\ref{app:dreff}, its orientation then diffuses with
\begin{equation}
D_r^{\mathrm{eff}}\simeq D_r+\frac{\langle\omega^2\rangle a}{2v}=D_r+\frac{\omega_0^2a}{6v}.
\label{eq:Dreff}
\end{equation}
The persistence length $v_0/D_r^{\mathrm{eff}}$ is reduced, and so is the effective P\'eclet number
\begin{equation}
\Pe_{\mathrm{eff}}=\Pe\,\frac{D_r}{D_r^{\mathrm{eff}}}.
\label{eq:Peeff}
\end{equation}

Table~\ref{tab:peeff} lists these estimates for $v=v_0$ and $a=\sigma$. They show that the torques we study are strong in the sense that matters. For $\Tx=15$ the effective P\'eclet number is already reduced from $138$ to $80$. For $\Tx=25$-$30$, that is $\Tx/v_0\approx0.2$, it is $35$-$46$, close to the critical value $\Pe_c\simeq40$~\cite{Siebert2018}. For $\Tx=100$ it is about $4$, far below. This is fully consistent with the kinetics. The activated regime appears for $\Tx=15$-$25$, $\ell(t)$ saturates for $\Tx=30$, and for $\Tx=100$ the system hardly phase-separates at all (Figs.~\ref{fig:time_evolution_torq} and \ref{fig:torq_length}). Because particles in crowded regions move more slowly than $v_0$, Eq.~\eqref{eq:Dreff} is a lower bound on the actual $D_r^{\mathrm{eff}}$. The measured long-time MSAD [Fig.~\ref{fig:torq_MSD_MSAD}(b)] is indeed somewhat larger than the estimate, by less than a factor of two.
% AUTHORS: please confirm the "less than a factor of two" from the MSAD data.

\begin{table}[t]
\caption{Random torques: maximal angular velocity $\omega_0=D_r\Tx/k_BT$, mean-square angular velocity, effective rotational diffusion from Eq.~\eqref{eq:Dreff} with $v=v_0$ and $a=\sigma$, effective P\'eclet number, Eq.~\eqref{eq:Peeff}, and the regime seen in Fig.~\ref{fig:torq_length}. For reference, $D_r=2.45$, $\Pe=138$, $\Pe_c\simeq40$.}
\label{tab:peeff}
\begin{ruledtabular}
\begin{tabular}{ccccccc}
$\Tx$ & $\Tx/v_0$ & $\omega_0$ & $\langle\omega^2\rangle$ & $D_r^{\mathrm{eff}}$ & $\Pe_{\mathrm{eff}}$ & regime\\
\hline
15 & 0.12 & 36.7 & 450 & 4.25 & 80 & activated\\
20 & 0.16 & 49.0 & 799 & 5.65 & 60 & activated\\
30 & 0.24 & 73.5 & 1799 & 9.64 & 35 & arrested\\
100 & 0.80 & 244.9 & 19986 & 82.4 & 4 & arrested\\
\end{tabular}
\end{ruledtabular}
\end{table}

This explains the large difference in efficiency between the two kinds of disorder. Forces must compete with the swim speed and become effective only when $\Fx\gtrsim v_0$. Torques compete with the much weaker rotational noise, and even a small torque, when accumulated over the many cells that a fast particle crosses per persistence time, dominates the orientational dynamics. A useful way to see this is to compare the rotation acquired in a single cell, $\omega_0a/v_0$, with the thermal rotation in the same time, $\sqrt{2D_ra/v_0}$. Their ratio is already about $1.5$ for $\Tx=15$ and about $12$ for $\Tx=100$.

\subsection{Why the spatial variation matters}

A uniform increase of the rotational diffusion coefficient would lower the P\'eclet number, and a uniform drag would lower the swim speed. Neither would change the universality class of coarsening. As long as the system remains above the critical point, it would still show LS growth, just with a different prefactor. The activated kinetics must therefore come from the \emph{spatial} variation of the disorder, which is what allows it to pin interfaces.

For torques, this spatial variation acts like a random local distance to the critical point. Regions where the local torques happen to be weak are effectively more active and favor the dense phase. Regions where they are strong favor the gas. This is the active analogue of random-bond (or random-$T_c$) disorder in a magnet, which also pins domain walls and leads to activated growth~\cite{Huse1985,Paul2004}. Dense domains survive where the torques are weak and must invade regions where they are strong. For forces, as discussed above, the spatial variation acts as a random field. In both cases, growing domains must cross an increasing number of unfavorable regions, which produces barriers that grow with domain size, as encoded in Eq.~\eqref{eq:barrier}.

A direct test of this picture would be to compare quenched disorder with annealed disorder of the same strength, i.e., fields redrawn at every time step. Our argument predicts that annealed disorder would restore $t^{1/3}$ growth, with a reduced prefactor, as long as $\Pe_{\mathrm{eff}}>\Pe_c$.

%===========================================================
\section{Discussion}
\label{sec:discussion}
%===========================================================

\emph{Two routes, one kinetic outcome.} The two kinds of disorder studied here could hardly be more different. One acts on where particles can go, the other on how long they keep going in one direction. Random forces behave like a random field for the density, whereas random torques behave more like random bonds, a spatially varying distance to the critical point. Yet both lead to the same kinetic behavior, described by the same scaling form, Eq.~\eqref{eq:zeff_scaling}, and the same activated, logarithmic growth law, Eq.~\eqref{eq:log_growth}. This suggests that activated coarsening is a generic consequence of quenched disorder in MIPS rather than a detail of how the disorder is introduced. The persistent energy input of the particles does not anneal the disorder away. Instead, it provides the noise that drives rare hops over barriers that grow with domain size.

\emph{Relation to earlier work.} Our results complement the steady-state analysis of Ro \emph{et al.}~\cite{Ro2021}, who showed that quenched random potentials and torques destroy bulk MIPS in two dimensions beyond an Imry-Ma length and produce long-range correlations. Our data show how the system gets there. Growth passes from a transient power law to a regime controlled by growing barriers, in the same way as domain growth in random magnets~\cite{Huse1985,Paul2004,Corberi2012}. The loss of the conserved-order-parameter dip in $C(r,t)$ is consistent with the long-range correlations they predicted, and the crossover length $\lambda$ decreases with disorder, as an Imry-Ma length should. The breakdown of superuniversality places active MIPS in the same class as the random-field Ising model, for which it was also reported~\cite{Corberi2012}.

\emph{Experimental implications.} Several of our findings can be tested directly. The effective exponent $\zeff(\ell)$ is easy to extract from time-lapse images of phase-separating active colloids and provides a sensitive diagnostic of activated growth. Speckle light fields~\cite{Volpe2014,Pince2016} produce tunable random forces on colloids, and spatially patterned illumination can produce spatially varying activity or torques~\cite{Lozano2016}. Bacteria in porous media~\cite{Bhattacharjee2019} provide a natural setting where disorder strength and swim speed can be varied. Our results make two concrete predictions for such experiments. First, disorder that scrambles orientations should arrest phase separation at a much weaker amplitude than disorder that pushes particles around. Second, force disorder should become effective roughly when the local drift matches the swim speed.

%===========================================================
\section{Conclusions}
\label{sec:conclusion}
%===========================================================

We have studied the coarsening of motility-induced phase separation of active Brownian particles in quenched random force and torque fields. In the clean system, domains grow according to the Lifshitz-Slyozov law $\ell\sim t^{1/3}$, with compact domains, sharp interfaces, and the correlation function and structure factor expected for a conserved order parameter. Quenched disorder, in either form, changes this qualitatively.

Coarsening remains self-similar at fixed disorder, but the scaling functions depend on disorder strength, so superuniversality fails. The conserved-order-parameter dip in the correlation function disappears, the Porod law breaks down, and interfaces become rough and pinned. The growth kinetics cross over from a transient power law with a disorder-dependent exponent to activated growth. This crossover follows the scaling form known from the random-field Ising model, and it implies barriers that grow with domain size as $\ell^{\varphi}$ and logarithmically slow growth, $\ell\sim\lambda(\ln t)^{1/\varphi}$, with $\phi\approx 2.2$ for random forces and $\phi \approx 1.3$ for random torques, and a crossover length $\lambda$ that decreases with disorder. 

The microscopic routes are different for the two kinds of disorder. Random forces leave orientations untouched but reduce translational mobility and, once the local drift exceeds the swim speed, trap particles in the sinks of a non-conservative random drift. They act as a random field on the density. Random torques act as a quenched, sign-random chirality. They make orientations spin ballistically where particles are stuck and diffuse faster where they move, so that the local P\'eclet number drops toward its critical value. Simple, parameter-free estimates reproduce the angular dynamics and explain why torques are several times more effective than forces.

These results show that quenched disorder is a relevant perturbation for the kinetics of active phase separation. They link MIPS coarsening to the physics of pinned interfaces in disordered magnets, and they suggest that in realistic, heterogeneous environments the size of active clusters is set by the disorder rather than by the age of the system.

% \begin{acknowledgments}
% AUTHORS: add funding information and acknowledgments.
% \end{acknowledgments}

%===========================================================
\appendix
%===========================================================

\section{Effective rotational dynamics in a random torque field}
\label{app:dreff}

In this Appendix we derive Eqs.~\eqref{eq:msad_short} and \eqref{eq:Dreff}. Consider a particle whose orientation obeys Eq.~\eqref{eq:eom_theta}, $\dot\theta=\omega(\mathbf r(t))+\sqrt{2D_r}\,\eta(t)$, where $\omega(\mathbf r)=D_r\Gamma(\mathbf r)/k_BT$ is constant within each lattice cell and independent from cell to cell, with $\langle\omega\rangle=0$ and $\langle\omega^2\rangle=\omega_0^2/3$.

\emph{Short times.} If the particle stays in one cell during the time $t$, then $\Delta\theta=\omega t+\sqrt{2D_r}\,W(t)$, where $W$ is a Wiener process independent of $\omega$. Averaging over noise and over cells,
\begin{equation}
\langle\Delta\theta^2\rangle=\langle\omega^2\rangle t^2+2D_rt=\frac{\omega_0^2}{3}t^2+2D_rt.
\end{equation}
The deterministic term dominates for $t>6D_r/\omega_0^2$, which is $t>1.1\times10^{-2}$ for $\Tx=15$ and $t>2.5\times10^{-4}$ for $\Tx=100$.

\emph{Long times.} A particle moving with speed $v$ crosses a cell of size $a$ in the time $\delta t=a/v$ and acquires a deterministic rotation $\delta\theta_j=\omega_j\,a/v$ in cell $j$. Because the $\omega_j$ are independent with zero mean, after $n=t/\delta t$ cells the accumulated deterministic rotation has variance
\begin{equation}
\left\langle\Big(\sum_{j=1}^{n}\delta\theta_j\Big)^2\right\rangle=n\,\langle\omega^2\rangle\frac{a^2}{v^2}=\langle\omega^2\rangle\frac{a}{v}\,t.
\end{equation}
This is diffusive growth with a coefficient $\langle\omega^2\rangle a/(2v)$. Adding the independent thermal contribution $2D_rt$ gives Eq.~\eqref{eq:Dreff}. The argument assumes straight motion across each cell. For a particle that slows down, $v$ is smaller and $D_r^{\mathrm{eff}}$ larger, so Eq.~\eqref{eq:Dreff} evaluated with $v=v_0$ gives a lower bound.

\section{From the effective exponent to the growth law}
\label{app:growth}

By definition, $d\ln t=\zeff\,d\ln\ell=\zeff\,d\ell/\ell$. Inserting Eq.~\eqref{eq:zeff_scaling} and integrating from a reference size $\ell_1$ to $\ell$ gives
\begin{equation}
\ln\frac{t}{t_1}=\bar z\ln\frac{\ell}{\ell_1}+b\int_{\ell_1}^{\ell}\left(\frac{\ell'-\ell_0}{\lambda}\right)^{\varphi}\frac{d\ell'}{\ell'}.
\end{equation}
For $\ell\gg\ell_0$ the integrand is approximately $(\ell'/\lambda)^{\varphi}/\ell'$, and the integral is $(1/\varphi)(\ell/\lambda)^{\varphi}$ up to terms that are negligible at large $\ell$. Collecting constants into $t_0$ yields
\begin{equation}
\ln\frac{t}{t_0}\simeq\bar z\ln\ell+\frac{b}{\varphi}\left(\frac{\ell}{\lambda}\right)^{\varphi},
\end{equation}
which is Eq.~\eqref{eq:arrhenius}. For $\ell\gg\lambda$ the second term dominates, and solving for $\ell$ gives Eq.~\eqref{eq:log_growth}. The crossover between the two regimes occurs where the two terms are comparable, $(\ell/\lambda)^{\varphi}\sim(\varphi\bar z/b)\ln\ell$.

\subsection*{Acknowledgements} 
B. Sen Gupta acknowledges Science and Engineering Research Board (SERB), Department of Science and Technology (DST), Government of India (no. CRG/2022/009343) for financial support. Parameshwaran A. acknowledges DST-SERB, India for doctoral fellowship.

\bibliography{refs}

\end{document}